\documentclass[a4paper,12pt,english,german]{article}

\begin{document}

\centerline{\bf Delayed choice without choice}

\centerline{\it M. Dugi\' c}

\centerline{Department of Physics, Faculty of Science, Kragujevac,
Serbia}

\bigskip

{\bf Abstract} A critical note on some of the existing proposals
for performing the "delayed choice" experiment is placed. By
abandoning the original idea and intention, some modern
theoretical proposals and experimental evidence are simply
incorrectly understood/interpreted. In effect, the Complementarity
principle remains practically intact.

\bigskip

There is ongoing debate on the possibility to perform the famous
Wheeler's "delayed choice experiment". However,  some proposals
 eg. [1] abandon the original spirit of the Wheeler's experiment
and necessarily draw incorrect conclusions (but see [2]).

The original proposal [3] is fascinating as it anticipates
something that we should  have already learned: {\it typically,
quantum systems do not have individual characteristics}.
Forgetting about the simple and elegant original intention, the
modern variations simply miss the point.

E.g., the following  state of the combined system "beam splitter +
photon (BS+P)":

\begin{equation}
\vert \psi \rangle = \sqrt{c} \vert 0\rangle_{BS} \vert particle
\rangle_P + \sqrt{1-c} \vert 1\rangle_{BS} \vert wave \rangle_P.
\end{equation}

In eq.(1):  $\vert particle \rangle_P$ represents the state of the
photon where it behaves like a particle and $\vert wave\rangle_P$
represents its state when it behaves like a wave. Eq.(1) is then
interpreted to provide the testing of both the wave and the
particle behavior of the photon at the {\it same time} [2, 4].

But this claim is in no sense different from, of course incorrect
[5], claiming that we have simultaneously measured both the
spin-up and the spin-down in the situation described in full
analogy with eq.(1) by:

\begin{equation}
\vert \phi \rangle = \sqrt{c} \vert spin-up\rangle_{spin} \vert +
\rangle_{apparatus} + \sqrt{1-c} \vert spin-down\rangle_{spin}
\vert - \rangle_{apparatus}.
\end{equation}

Quantum {\it sub}systems are open systems not possessing states of
their own [5]. Without a definite initial state, the photon in
eq.(1) is deprived of  individuality and therefore cannot make any
"choice".

\bigskip

[1] R. Ionicioiu, D. R. Terno, Phys. Rev. Lett. 107, 230406 (2011)

[2] T. Kureshi, arXiv:1205.2207 [quant-ph]

[3] J. A. Wheeler, pp. 9-48 in Mathematical Foundations of Quantum
Mechanics, edited by A.R. Marlow (Academic, New-York, 1978).

[4] A. Peruzzo et al, arXiv:1205.4926 [quant-ph]

[5] B. d' Espagnat, Conceptual Foundations of Quantum Mechanics,
Reading, 1976 [second edition]

\end{document}